\documentclass[a4paper,aps,pra,twocolumn,floatfix,superscriptaddress,notitlepage,usenames,dvipsnames,svgnames,table,footinbib,longbibliography,nofootinbib]{revtex4-1}
\usepackage[english]{babel}
\usepackage{letltxmacro}
\LetLtxMacro{\ORIGselectlanguage}{\selectlanguage}
\usepackage[utf8]{inputenc}
\newtheorem{theorem}{Theorem}
\usepackage[margin= 0.75in]{geometry}
\usepackage{fullpage,amssymb,amsmath,xcolor,cancel,gensymb,hyperref,graphicx,mathtools}
\usepackage{advdate}
\usepackage{indentfirst}
\usepackage{physics}
\usepackage[cal=cm]{mathalfa}
\usepackage{graphicx}
\graphicspath{ {./images/} }

\DeclareFontFamily{OT1}{pzc}{}
\DeclareFontShape{OT1}{pzc}{m}{it}{<-> s * [1.10] pzcmi7t}{}
\DeclareMathAlphabet{\mathpzc}{OT1}{pzc}{m}{it}
\usepackage{afterpage}

\hypersetup{colorlinks,linkcolor=blue,citecolor=red,urlcolor=cyan}

\usepackage[breakable]{tcolorbox}
\usepackage{parskip} 
    
\usepackage{iftex}
\ifPDFTeX
	\usepackage[T1]{fontenc}
    	\usepackage{mathpazo}
    \else
    	\usepackage{fontspec}
    \fi

    \usepackage{graphicx}
    
    \usepackage{caption}
    \DeclareCaptionFormat{nocaption}{}
    \usepackage{float}
    \usepackage{xcolor} 
    \usepackage{enumerate} 
    \usepackage{geometry} 
    \usepackage{amsmath} 
    \usepackage{amssymb} 
    \usepackage{textcomp} 
    \AtBeginDocument{%
    }
    \usepackage{upquote} 
    \usepackage{eurosym} 
    \usepackage[mathletters]{ucs} 
    \usepackage{fancyvrb} 
    \usepackage{grffile} 
    \makeatletter 
    \@ifpackagelater{grffile}{2019/11/01}
    {
    }
    {
      \def\Gread@@xetex#1{%
        \IfFileExists{"\Gin@base".bb}%
        {\Gread@eps{\Gin@base.bb}}%
        {\Gread@@xetex@aux#1}%
      }
    }
    \makeatother
    \usepackage[Export]{adjustbox} 
    \adjustboxset{max size={0.9\linewidth}{0.9\paperheight}}

\usepackage[font=small,labelfont=bf,justification=RaggedRight,format=plain]{caption}
\usepackage{subcaption}
\usepackage{cleveref}

\begin{document}
\title{No $((n,K,d< 127))$ code can violate the quantum Hamming bound}

\author{Emanuel Dallas}
\email [e-mail: ]{dallas@usc.edu}

\author{Faidon Andreadakis}
\email [e-mail: ]{fandread@usc.edu}

\affiliation{Department of Physics and Astronomy, and Center for Quantum Information Science and Technology, University of Southern California, Los Angeles, California 90089-0484, USA}

\author{Daniel Lidar}
\email [e-mail: ]{lidar@usc.edu}

\affiliation{Department of Physics and Astronomy, and Center for Quantum Information Science and Technology, University of Southern California, Los Angeles, California 90089-0484, USA}
\affiliation{Department of Electrical \& Computer Engineering, University of Southern California, Los Angeles, California
90089, USA}
\affiliation{Department of Chemistry, University of Southern California, Los Angeles, California 90089, USA}

\date{\today}

\begin{abstract}
It is well-known that non-degenerate quantum error correcting codes (QECCs) are constrained by a quantum version of the Hamming bound. Whether degenerate codes also obey such a bound, however, remains a long-standing question with practical implications for the efficacy of QECCs. We employ a combination of previously derived bounds on QECCs to demonstrate that a subset of all codes must obey the quantum Hamming bound. Specifically, we combine an analytical bound due to Rains with a numerical bound due to Li and Xing to show that no $((n,K,d))$ code with $d< 127$ can violate the quantum Hamming bound.
\end{abstract}

\maketitle

\section{Introduction} \label{secint}

The field of quantum computation has recently exploded with a wide range of exciting results regarding the practical applicability of quantum algorithms and the ability to realize ever-growing quantum processor sizes~\cite{preskill_quantum_2018}. 
Naturally, meaningful large-scale realizations of quantum algorithms depend on the ability to achieve fault tolerant computation~\cite{Campbell:2017aa}.

Unfortunately for this endeavor, it is extremely difficult to implement quantum circuits with fidelities large enough to achieve useful computational tasks \cite{s2020compilation}. Environmental noise and imperfect physical implementations of unitary gates lead to errors, which tend to propagate and grow throughout circuits~\cite{Suter:2016aa}.

Various methods have been developed to render functional and scaled up quantum circuitry possible in spite of this noise~\cite{lidar}.
Error correction requires redundantly encoding information to ensure that it is fully recoverable after being damaged by some particular set of errors. 
Quantum error correction is a general and robust method for reducing the effects of noise~\cite{shor_scheme_1995,Steane:96a}.
An $((n,K,d))$ quantum error-correcting code (QECC) is a $K=q^k$ dimensional subspace of the Hilbert space of $n$ $q$-level particles that is characterized by its distance $d$, the largest integer such that the code can detect any error that acts non-trivially on at most $d-1$ such particles~\cite{Calderbank:96}. The most common case is $q=2$, where each particle is a qubit.
In recent years, there has been significant progress in experimentally realizing many-qubit quantum error correction \cite{chiaverini_realization_2004,2012Natur.482..382R,andersen_repeated_2020,ryan-anderson_realization_2021,google_quantum_ai_exponential_2021,krinner_realizing_2022,marques_logical-qubit_2022}.

For a given quantum information processing task, it would be desirable to maximize both the rate $k/n$ and the distance $d$ of the QECC one selects. In order to guide these efforts, it is imperative to have an understanding of the space of possible codes in terms of the parameters $((n,K,d))$. Since the advent of quantum error correction, there has existed an active field of research devoted to obtaining bounds on achievable codes for the parameter space of $((n,K,d))$~\cite{Grassl:codetables}. Such bounds quantify the inevitable tradeoff between the rate and the distance.

The quantum Hamming bound (QHB) is one such bound known to be satisfied by non-degenerate codes, which are codes for which linearly independent errors map the codespace to linearly independent subspaces~\cite{Ekert:1996}. It is currently unknown for what set of degenerate codes the QHB holds.
Some results have shown violations of Hamming-like bounds for various codes~\cite{cats,shor_quantum_1996,kay,subsystemcodes,baconshor,cliff} (which do not immediately bear a clear resemblance to the QHB), but others conjecture that the QHB holds for all quantum codes \cite{degen,calderbank}. Various methods have been employed to calculate constraints on $n$ and $d$ values for which the QHB may be violated \cite{rains, degen, lixing, yu}. 
Degenerate QHB-violating codes could be of great practical importance since they would more efficiently correct errors (achieving the same $k,d$ for a lower $n$ ``cost''). However, research on the topic has progressed little in recent years. Most of the results have been negative, reducing the range of $((n,K,d))$ values for which codes may exist that would violate the QHB.
In this work we contribute to this effort by combining an analytical bound derived by Rains~\cite{rains} with novel calculations of a numerical bound derived by Li and Xing~\cite{lixing}, to show that no $((n,K,d))$ code can violate the QHB for $d<127$.

In \autoref{secpre}, we introduce the statement of the QHB and describe the intuition behind its derivation. In \autoref{secbound}, we introduce the key bounds we employ in this paper. In \autoref{secres}, we combine these bounds to derive the impossibility of QHB-violating codes of distance $d < 127$. Finally, we outline possible extensions of this work.

\section{Preliminaries} \label{secpre}

Let
\begin{equation}
\mathcal{H}\cong \mathcal{H}_C \oplus \mathcal{H}_\perp \cong \mathbb{C}^{q^k} \oplus \mathbb{C}^{q^{n-k}}
\label{eq:1}
\end{equation}
be the Hilbert space associated with the quantum system. Here, $n$ corresponds to the number of physical $q$-level particles and the codespace $\mathcal{H}_C$ comprises $k$ encoded particles. Vectors in $\mathcal{H}_C$ correspond to code states. The state $\rho$ of the quantum system is a positive semidefinite, unit trace element of the space of bounded linear operators acting on $\mathcal{H}$ ($\rho \geq 0$, $\Tr[\rho ]=1$, $\rho\in \mathcal{B}(\mathcal{H})$).

Henceforth, unless otherwise noted, we consider only binary codes in this work, i.e., codes defined over qubits ($q=2$) as opposed to $q$-dimensional systems with $q>2$. Much of the theoretical background we present is easily generalizable to the higher-dimensional setting (e.g., see \cref{sec:LXB}), but our numerical results hold only for $q=2$.

For an arbitrary quantum circuit, let $\mathcal{N}$ be the completely positive, trace-preserving (CPTP) map representing the noise process. We may write $\mathcal{N}$ in the Kraus operator sum representation (OSR)~\cite{K83} as
\begin{equation}
    \mathcal{N}(\rho) = \sum_j E_j \rho E_j^{\dagger},
\end{equation}
where $\{E_j\}$ is a set of linearly independent ``error operators'' corresponding to different possible errors acting on the system of qubits~\cite{knill:1997kx}. For a system of $n$ qubits, we may represent the $E_j$ in the $n$-qubit Pauli basis, spanned by operators of the form $\prod_{m=1}^n \sigma_m^{i_m}$, where $i_m \in \{x,y,z,0\}$ and $\sigma^0\equiv I$, the identity operator. An error of weight $w$ is one in which $i_m \neq 0$ on strictly $w$ of the $n$ qubits. A detectable error is an error $E_j$ which maps every state in $\mathcal{H}_C$ to a state in $\mathcal{H}_\perp$. A correctable error is a detectable error which can be ``undone'' without knowing or destroying the quantum superposition state. Formally, a set of errors $\{E_j\}$ is correctable by a code $C$ iff it obeys the Knill-Laflamme condition~\cite{knill:1997kx}:
\begin{equation}
    P_C E_i^{\dagger}E_j P_C = \gamma_{ij}P_C \quad \forall i,j,
    \label{eq:KL} 
\end{equation}
where $\gamma$ is a Hermitian matrix. Here $P_C = \sum_{\alpha=1}^K \ket{{\bar{\psi}_\alpha}}\!\bra{{\bar{\psi}_\alpha}}$ is the codespace projector from $\mathcal{H}$ to $\mathcal{H}_C$, and the set $\{\ket{\bar{\psi}_\alpha}\}$ forms an orthonormal basis for $\mathcal{H}_C$. Eq.~\eqref{eq:KL} means that orthogonal code states remain orthogonal after the noise map, i.e., the encoded information is not corrupted. The condition for a detectable set of errors replaces Eq.~\eqref{eq:KL} by $P_C E_j P_C = \gamma_{j}P_C$ $\forall E_j$~\cite{knill:1997kx}.

Consider a \emph{non-degenerate} code~\cite{calderbank}: a code for which different elements of the set of correctable errors map any code state to linearly independent states:
\begin{equation}
E_i \ket{\bar{\psi}_\alpha} \neq E_j \ket{\bar{\psi}_\alpha} \; \forall \alpha , \, i\neq j.
\end{equation}
Non-degeneracy is equivalent to $\gamma$ in Eq.~\eqref{eq:KL} having full rank.

Since the number of linearly independent vectors in a Hilbert space is bounded by the dimension of the space itself, \emph{non-degenerate} $((n,K,d))$ binary quantum codes obey the quantum Hamming bound~\cite{Ekert:1996,calderbank}:
\begin{equation}
\sum_{j=0}^{t}\binom{n}{j}3^j  \leq 2^{n-k} ,
\label{ham_bound}
\end{equation}
where $t=\lfloor (d-1)/2 \rfloor$ is the highest error weight for which the code can correct all errors. The QHB can be derived from a simple combinatorial counting argument: it states that the number of linearly independent (LI) states with no errors ($2^k$) plus the number of LI states with correctable errors ($2^k\sum_{j=1}^{t}\binom{n}{j}3^j$) cannot exceed the total number of LI states in the Hilbert space ($2^n$).

On the other hand, consider a \emph{degenerate} code. For such a code, distinct correctable errors may have the same effect on the same state. This means that there exist some $i, j$ such that:
\begin{equation}
    E_i\ket{\bar{\psi}_\alpha} = E_j\ket{\bar{\psi}_\alpha} \forall \alpha .
\end{equation}
Degeneracy is equivalent to $\gamma$ in Eq.~\eqref{eq:KL} being rank-deficient.

Since in the degenerate case the vectors resulting from the action of different errors need \emph{not} be linearly independent, the number of correctable errors may exceed the bound set by the dimension of the Hilbert space. This leaves open the possibility that a degenerate code may violate the quantum Hamming bound.\footnote{There exists a related but different dichotomy, between pure and impure codes. Whereas non-degeneracy is concerned with linear independence, purity is about orthogonality. More precisely, if different errors $E_i$ and $E_j$ map $\mathcal{H}_C$ to mutually orthogonal subspaces, we
say that $C$ is pure with respect to $\{E_j\}$. A pure code must be non-degenerate. For the important class of additive (or stablizer~\cite{gottesman}) codes, i.e., quantum analogues to additive classical codes, derived from orthogonal geometry, ``non-degenerate'' and ``pure'' are
equivalent~\cite{calderbank}. In general, a pure code is non-degenerate but the converse need not be true~\cite{calderbank}. Indeed, there exists a non-additive $((7,2,3))$ code that is non-degenerate but impure~\cite{Cao2022quantumvariational}.} 

It is important to note that code degeneracy and impurity are strictly quantum phenomena. Since any classical error is characterized by which bits it flips, all distinct classical errors map a given code state to distinct states. However, in quantum codes, the possibility of phase flips in conjunction with the Hilbert space structure of the code space, allows distinct errors to map to identical states. Interestingly, the first quantum error correcting code ever discovered was a degenerate code~\cite{shor_scheme_1995}.

\section{Bounds on quantum codes}
\label{secbound}

We make use of two bounds in this work, which we review in this section. The first is the Shadow Enumerator Bound discovered by Rains, that sets a limit on how many errors an arbitrary $n$-qubit code can correct~\cite{rains}.
The second is a bound due to Li and Xing that allows us to calculate an integer $N>0$, given a distance $d$, for which all $((n,K,d))$ codes with $n>N$ satisfy the QHB~\cite{lixing}.

\subsection{Rains bound}

In \cite[Theorem~15]{rains}, Rains used shadow enumerator techniques to prove the following:
\begin{theorem}
If a ((6m-1+l,K,d)) quantum error-correcting code exists for $K>1$ with $0\leq l \leq 5$, then
\begin{equation}
    d\leq \begin{cases} 2m+1 & \text{if } l<5 \\ 2m+2 & \text{if } l=5 \end{cases} .
\end{equation}
\end{theorem}
From the above theorem, by setting $n=6m-1+l$, we have $d \leq \frac{n+4-l}{3} \leq \frac{n+4}{3}$ for $0\leq l<5$ and $d \leq \frac{n+2}{3} < \frac{n+4}{3}$ for $l=5$, i.e., for any code with $n$ physical qubits we obtain the bound (which we refer to as the Rains bound):
\begin{equation} 
    n\geq 3d-4. 
    \label{shadow}
\end{equation}

\begin{figure*}
\includegraphics[]{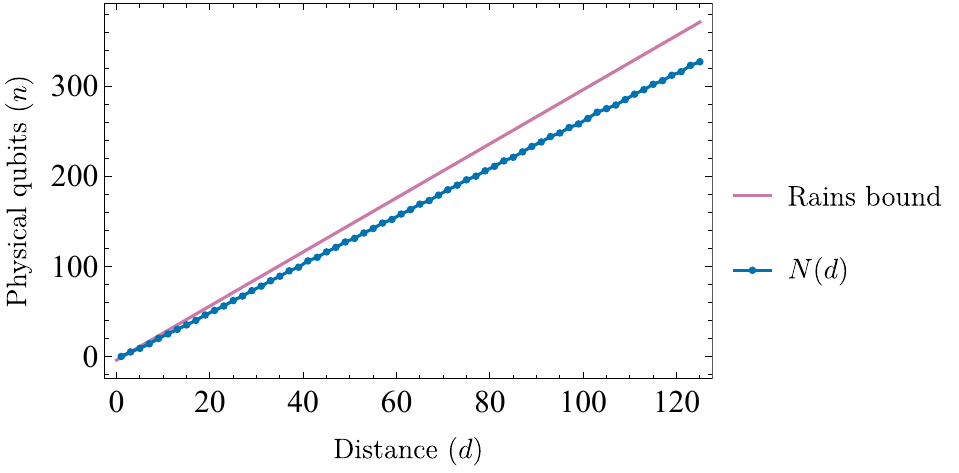}
\caption{The distance versus number of physical qubits, for both the Rains bound~\cref{shadow} and the Li-Xing quantity $N(d)$~\cref{condition}. All codes that satisfy the Li-Xing bound $n>N(d)$ also satisfy the QHB. But every code must also satisfy the Rains bound, i.e., lie above the purple line. Therefore, for $d$ values where the Rains bound is above $N(d)$, no codes of that distance can violate the QHB. Conversely, if there exists a $d$ value where the Rains bound falls below $N(d)$ then this would imply the existence of a code that can potentially violate the QHB; we have found no evidence for this up to $d< 127$.}
\label{fig}
\end{figure*}

\subsection{Li-Xing bound}
\label{sec:LXB}

Li and Xing have shown that there exists an integer $N(d, q)$ such that any $((n,K,d))_q$ code for which $n>N$ must satisfy the QHB \cite{lixing}, an idea posited earlier by Aly \cite{alyproofsketch}. Here $q$ again refers to a $q$-level quantum system. As mentioned above, we assume we are dealing with qubits ($q=2$) for simplicity, but the results can be generalized to $q>2$.

Let $P_i(x;n)$ denote the $i$-th Krawtchouk polynomial, defined as follows: \begin{equation}
P_i(x;n) \equiv \sum_{j=0}^i (-1)^j \, 3^{i-j} \binom{x}{j} \binom{n-x}{i-j}, \; i=0,1,\dots
\end{equation}

Here, the $3$ comes from $q^2 - 1$, where we have used the above specification of $q=2$. Krawtchouk polynomials are ubiquitous in both classical~\cite{macwilliams1977theory} and quantum coding theory~\cite{PhysRevLett.78.1600}. The following theorem was proved in Ref.~\cite{lixing} (which follows from a result in Ref.~\cite{oldlixing}).

\begin{theorem} \label{thm2}
Given an $((n,K,d))$ code, define S to be a non-empty subset of $\{0,...,d-1\}$ and define $T \equiv \{0,...,n\}$. If the function $f(x;n) \equiv \sum_{i=0}^n f_{i;n} \, P_i(x;n)$ satisfies

\begin{enumerate}
\item $f_{g;n} > 0 \ \forall g \in S, \ f_{g;n} \geq 0$ otherwise.
\item $f(g;n) \leq 0 \ \forall g \in T \setminus S$
\end{enumerate}

then $2^k \leq \frac{1}{2^n}\max\limits_{g\in S} \frac{f(g;n)}{f_{g;n}}$.
\end{theorem}

Select $f_{g;n} \equiv \big( \sum_{i=0}^t P_i(g;n)\big)^2$. Then~\cite{lixing}:
\begin{equation}
    \frac{f(0;n)}{f_{0;n}} = \frac{2^{2n}}{\sum_{j=0}^t 3^j \, \binom{n}{j}}.
\end{equation}
Let us now choose $S \equiv \{ 0, \dots, d-1 \}$. Considering large $n$ expansions, Ref.~\cite{lixing} showed that
\begin{subequations}
\begin{align}
&f_{g;n} = (3^t/t!)n^{2t}+ o(n^{2t}) \quad \text{for } g\in S \\
&f(g;n) = 3^{2n} \, o(n^t) \quad \text{for } g\in S \\
&f(g;n)=0 \quad \text{for } g\in \{ d, \dots, n \}
\end{align}
\end{subequations}
and
\begin{equation}
f(0;n) = 3^{2n}\left((3^t/t!)n^t + o(n^t)\right).
\end{equation}
The asymptotic behavior implies that there exists an integer $N(d)$ such that $\forall n>N(d)$:\footnote{The same technique, with a different choice of $f_{t;n}$, was used in \cite{litsyn} to derive the quantum Hamming and Singleton bounds.}
\begin{equation}
\max\limits_{g\in S}\frac{f(g;n)}{f_{g;n}} = \frac{f(0;n)}{f_{0;n}} . \label{condition}
\end{equation}

Li and Xing do not explicitly prove that the minimum $n$ for which this condition holds is $N(d)$, but they take this to be true in their numerical calculations of $N(d)$. We conjecture this in our work as well, and checked the conjecture numerically as explained below.
Succinctly:
\begin{align}
\label{succinct}
&N(d) = \\
&\quad \min \left\{ n_d \in \mathbb{Z} \bigg\vert \arg\max\limits_{g \in S} \frac{f(g;n)}{f_{g;n}} = 0 \;\; \forall n>n_d\right\}. \notag
\end{align}

Then, \autoref{thm2} ensures that for such $n$
\begin{align}
    2^k \leq \frac{1}{2^n} \, \frac{2^{2n}}{\sum_{j=0}^t 3^j \binom{n}{j}} =  \frac{2^{n}}{\sum_{j=0}^t 3^j \binom{n}{j}},
\end{align}
which \emph{is} the QHB. In other words, by calculating $N(d)$, we are guaranteed that as long as $n>N(d)$, all QECCs with these $n,d$ values satisfy the QHB. Also, in Ref. \cite{lixing} it is shown that 
$N(2m) = N(2m+1)$ for all $m\ge 1$. This allows us to limit our calculations to odd $d$ only.

Equation~\eqref{succinct} provides us with a means to compute $N(d)$ numerically, and the result is shown as the blue dots in \autoref{fig}. More explicitly, we obtained the $N(d)$ values shown in \autoref{fig} as follows: first, we generate a list of binomial coefficients with non-negative arguments using the recursive relationship
\begin{equation}
    \binom{n}{k} = \binom{n-1}{k-1} + \binom{n-1}{k}.
\end{equation}
Next, for a given $d$, we loop over increasing values of $n$, starting with $n=1$. For each $n$, we calculate $\frac{f(g)}{f_g}$ $\forall g \in S$. If $\arg\max\limits_{g \in S} \frac{f(g)}{f_g} = 0$, we set $N(d) = n$ and end the loop for that $d$ value, then move on to $d+2$. Otherwise, we continue looping. Our method relies on the minimum $n$ for which the condition $\eqref{condition}$ is satisfied to coincide with $N(d)$. To gain confidence that such a hypothesis may have merit, we selected $d=21$ and checked numerically that the condition \eqref{condition} indeed holds $\forall n \in \{51,499\}$ (for $n>499$ the numerical computations became unfeasible). 

\begin{figure}
\includegraphics[scale=1.0]{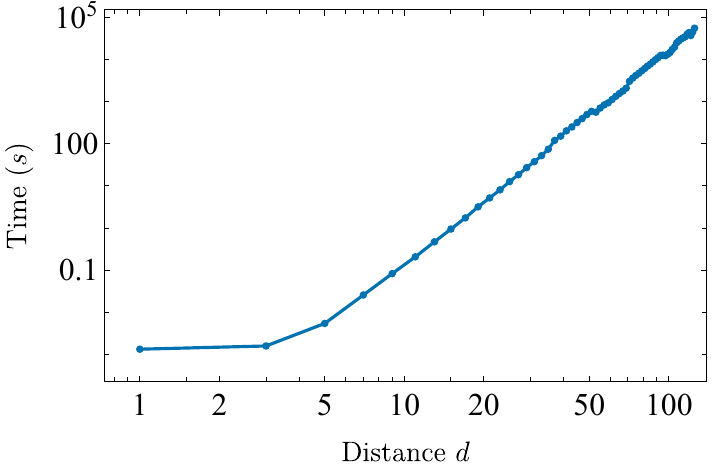}
\caption{Log-log plot of the time required to search for $N(d)$ as a function of the distance $d$. We observe that the time scaling is polynomial: the best linear fit in the log-log plane starting from $d=5$ gives a slope of $5.089$ with correlation coefficient $0.9995$. For reference, the two extreme data points are $(d=5,T=0.005412579$ s$)$ and $(d=125,T=55732.4488$ s$)$. For some values of $d$ there are some small but discernible drops in the computation time; we attribute such behavior to our use of dynamic programming methods: subroutines used for smaller $d$ are stored and re-utilized for larger $d$, which can sometimes reduce the computation time as $d$ is increased.}
\label{fig_time}
\end{figure}

The efficiency of our numerical method allowed us to compute $N(d)$ for up to $d=125$, thereby demonstrating non-existence of QHB-violating codes for $d<127$. Data collection required one week on a high-performance computing cluster at the University of Southern California's Center for Advanced
Research Computing (CARC). The scaling of time ($T$) required to numerically search for $N(d)$ is polynomial in $d$ with $T = O(d^5)$ (see \autoref{fig_time}). There are no known analytic shortcuts for calculating $N(d)$. Our Python code can be found in Ref.~\cite{andreadakis_qhb_2022}.

\section{Results and Discussion} 
\label{secres}

The Rains bound Eq.~\eqref{shadow} restricts the space of allowed QECCs by providing a lower bound on $n$. Simultaneously, for sufficiently large $n$ the Li-Xing bound ensures that the QHB is satisfied. Due to these observations, we numerically find $N(d)$ for odd $d$ up to $125$ by searching for the lowest $n$ for which Eq.~\eqref{condition} is satisfied, recovering and extending previous results by Li and Xing, who computed $N(d)$ up to $d=15$~\cite{lixing}. Then, in combination with the Rains bound, we rule out the existence of QHB-violating codes with particular distances. \autoref{fig} illustrates this point. From the Rains bound, the only possible codes are ones that lie \textit{above} the solid purple line. From the Li-Xing bound, we know that any code lying above the dot-dashed blue line \textit{must obey the QHB}. Therefore, if the Rains bound lies above the Li-Xing bound for a given distance $d$, all existing distance-$d$ codes must obey the QHB.

One may worry about the fact that the graphs cross at $d=1$. However, recall that 
the distance $d$ is the lowest weight of those errors which the code cannot detect, so a code with $d=1$ cannot detect (let alone correct) any error. 

There is a wide range of possible future work with the above bounds. First, and most directly, higher-powered computers can continually calculate the Li-Xing $N(d)$ for larger $d$ values; our empirical observation is that the time required scales as $\sim d^5$, so this calculation is relatively benign. If $N(d)$ exceeds the Rains bound for a particular $d$ value, it remains undetermined by this method whether or not there exists a QHB-violating code of distance $d$. In this case, one may employ known linear programming bounds to determine if a given $((n,K,d))$ code exists \cite{rains, gottesman, calderbank}.

Another approach would be to work with the Li-Xing bound analytically. To prove that the QHB holds for all QECCs, one must first rigorously prove the conjectured Eq.~\eqref{succinct}. It is then sufficient, but not necessary, to prove that the derivative of $N(d)$ is less than $3$ everywhere [recall Eq.~\eqref{shadow}]. While this holds for our dataset, it is not clear that this should hold for all $d$. One could also attempt to directly analytically demonstrate that $N(d)$ is less than the Rains bound for all $d$, regardless of its derivative. Both $N(d)$ and its derivative are extremely complicated functions, and either of these results would be truly remarkable.

\acknowledgments

Research was sponsored by the Army Research Office and was
accomplished under Grant Number W911NF-20-1-0075. The views and conclusions contained in this
document are those of the authors and should not be interpreted as representing the official policies, either
expressed or implied, of the Army Research Office or the U.S. Government. The U.S. Government is
authorized to reproduce and distribute reprints for Government purposes notwithstanding any copyright
notation herein.


%

\end{document}